\documentclass[aps,twocolumn,prx,preprintnumbers,showpacs,amsmath,amssymb,superscriptaddress]{revtex4-1}
\usepackage[dvipdfm, colorlinks, citecolor=blue]{hyperref} 
\usepackage{graphicx}
\usepackage{dcolumn}
\usepackage{threeparttable}
\usepackage{multirow}
\usepackage{booktabs}
\usepackage{txfonts}
\usepackage{xcolor}
\usepackage{bm}
\usepackage{amssymb}
\usepackage{amsmath}
\usepackage{latexsym}
\usepackage{epsfig}
\usepackage{amsbsy}
\usepackage{array}
\usepackage{tabularx}
\usepackage{esvect} 
\usepackage{extarrows}
\usepackage{float}

\newcommand{\rr}{\mathbf{r}}

\begin{document}

\title{$\alpha$-$\beta$ phase transition of zirconium predicted by on-the-fly machine-learned force field}

\author{Peitao Liu}
\email{peitao.liu@univie.ac.at}
\affiliation{VASP Software GmbH, Sensengasse 8, 1090 Vienna, Austria}

\author{Carla Verdi}
\affiliation{University of Vienna, Faculty of Physics and Center for
Computational Materials Science, Sensengasse 8, A-1090, Vienna,
Austria}

\author{Ferenc Karsai}
\affiliation{VASP Software GmbH, Sensengasse 8, 1090 Vienna, Austria}

\author{Georg Kresse}
\affiliation{University of Vienna, Faculty of Physics and Center for
Computational Materials Science, Sensengasse 8, A-1090, Vienna,
Austria}
\affiliation{VASP Software GmbH, Sensengasse 8, 1090 Vienna, Austria}

\begin{abstract}
The accurate prediction of solid-solid structural phase transitions at finite temperature is a challenging task,
since the dynamics is so slow that direct simulations of the phase transitions by
first-principles (FP) methods are typically not possible. Here, we study the $\alpha$-$\beta$ phase
transition of Zr at ambient pressure by means of on-the-fly machine-learned
force fields. These are automatically generated during FP molecular dynamics (MD)
simulations without the need of human intervention, while retaining almost FP accuracy.
Our MD simulations successfully reproduce the first-order displacive nature of
the phase transition, which is manifested by an abrupt jump of the volume
and a cooperative displacement of atoms at the phase transition temperature.
The phase transition is further identified by the simulated x-ray powder diffraction,
and the predicted phase transition temperature is in reasonable agreement with experiment.
Furthermore, we show that using a singular value decomposition and pseudo inversion of the design matrix
generally improves the machine-learned force field compared to the usual inversion
of the squared matrix in the regularized Bayesian regression.
\end{abstract}

\maketitle

\section{Introduction}\label{sec:intro}

Because of widespread applications in nuclear, chemical, and manufacturing
process industries~\cite{NORTHWOOD198558, KALAVATHI2019781},
zirconium has stimulated extensive interest in fundamental research
aiming to clarify the underlying mechanisms responsible for the phase transitions and phase diagram from both
experiment and theory~\cite{Vogel1931_T1136K,Fisher_Exp1964, Olinger1973,Sikka_review1982,
Xia_PRLExp1990, *Xia_PRB1991,Song_Exp1995,Zhao_EOS_Exp2005,ZHANG_exp2005,LiuWei_JAP_Exp2008,
Akahama_Exp1991, *Akahama_Exp1992, Ostanin_DFT1998,Stavrou_Exp2018,
BURGERS1934561, Petry_Exp1991, Ahuja_DFT1993, Greeff_DFT2005,Schnell_DFT2006, Souvatzis_DFT2011,
Chen_phonon_cal_PRB1985,Zong_ML_npj2018,Souvatzis_PRL2008,Qian_ML_PRB2018,
Stassis_phonon_PRL1978,Ye_phonon_cal_PRL1987,Heiming_phonon_Exp1991,HU_DFT2011,Willaime_MD_PRL1989,Zong_ML_PRB2020,PhysRevB.84.180301}.
Upon cooling the melt, Zr solidifies to a body-centred cubic (bcc) structure (the $\beta$ phase)
and undergoes a phase transformation to a hexagonal close-packed (hcp) structure (the $\alpha$ phase)
at a temperature lower than 1136 K at zero pressure~\cite{Vogel1931_T1136K}
and at lower temperatures under pressure~\cite{ZHANG_exp2005}.
With increasing pressure, the hcp phase transforms into another hexagonal but not close-packed structure
(the $\omega$ phase)~\cite{Olinger1973,Sikka_review1982,Xia_PRLExp1990,Song_Exp1995,Zhao_EOS_Exp2005,ZHANG_exp2005,LiuWei_JAP_Exp2008}.
Under further increased pressure, the $\omega$ phase transforms to the $\beta$ phase~\cite{Xia_PRLExp1990,ZHANG_exp2005}.
The experimentally estimated $\alpha$-$\omega$-$\beta$ triple point is at 4.9 GPa and 953 K~\cite{ZHANG_exp2005}.

\begin{figure}[H] 
\begin{center}
\includegraphics[width=0.48\textwidth,clip]{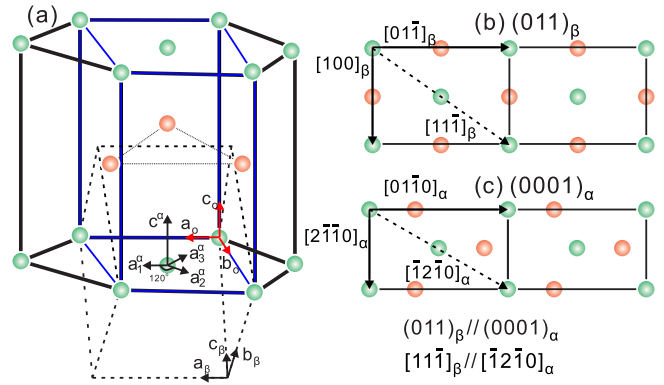}
\end{center}
 \caption{(a) Schematic illustration of the structural relationship between hcp ($\alpha$) and bcc ($\beta$) Zr. The black straight lines
 and dashed lines represent the hcp and bcc conventional unit cells, respectively. Note that for simplicity,
 atoms are only shown for the hcp phase in (a). The blue lines indicate the minimum common 4-atom orthorhombic (o) cell,
 whose lattice parameters ($a_o$, $b_o$, $c_o$) defined in terms of hcp and bcc lattices are given
 as ($a_1^\alpha$, $a_1^\alpha+2 a_2^\alpha, c^\alpha$) and ($a_\beta$, $b_\beta-c_\beta$, $b_\beta+c_\beta$), respectively.
 (b) The (011) plane of the bcc phase. (c) The (0001) plane of the hcp phase. Green and red balls represent the Zr atoms in two layers.
The crystallographic relation for the $\beta$-$\alpha$ martensitic phase transition is indicated~\cite{BURGERS1934561}.
}
\label{fig:str_plot}
\end{figure}

To understand the microscopic mechanism of the bcc-hcp phase transition of Zr,  Burgers~\cite{BURGERS1934561}
proposed that the transition can be divided into two processes.
As illustrated in Fig.~\ref{fig:str_plot}, the bcc phase first undergoes a long wavelength shear in the
[11$\bar{1}$] direction along the (112) plane (or equivalently in the  [1$\bar{1}$1] direction along the ($\bar{1}1$2) plane),
which squeezes the bcc octahedron to the hcp one, thereby changing the angle between the
[11$\bar{1}$] and  [1$\bar{1}$1] directions from 109.5$^{\circ}$ to 120$^{\circ}$~\cite{BURGERS1934561,Petry_Exp1991}.
Then, the neighboring (011) planes of the bcc phase experience a shuffle along opposite [01$\bar{1}$]
directions with a displacement of $a_\beta\sqrt{2}/12$~\cite{BURGERS1934561,Petry_Exp1991}  [compare Figs.~\ref{fig:str_plot}(b) and (c)].
The shuffle originates from displacements along the zone-boundary $N$-point
phonon of the $T_1$ branch in the [110] direction~\cite{BURGERS1934561,Petry_Exp1991}.
The transition belongs to the martensitic transformations, is of first order and displacive, and adopts the definite orientational
crystallographic relation (011)$_\beta$ //(0001)$_\alpha$ and [11$\bar{1}$]$_\beta$ //[$\bar{1}$2$\bar{1}$0]$_\alpha$~\cite{BURGERS1934561}.

The Burgers mechanism was later confirmed by Willaime and Massobrio~\cite{Willaime_MD_PRL1989} using
classic molecular-dynamics (MD) simulations based on a semi-empirical tight-binding interatomic potential~\cite{Willaime_PRB1991},
giving valuable insight on the temperature-induced hcp-bcc phase transition of Zr from an atomistic point of view.
However, their predicted phase transition temperature deviated by nearly 800 K from the experimental value,
since their potential was fitted to the hcp Zr phase only~\cite{Willaime_MD_PRL1989}.
By including zero-temperature as well as high-temperature properties of both hcp and bcc Zr phases
in the fitting procedure, Mendelev and Ackland~\cite{Mendelev_EAM_2007}
developed an embedded-atom interatomic potential that predicted a reasonable
hcp-bcc transition temperature. Some residual dependency on the target properties used in the fitting, however, remained.
Furthermore, these physics-based semi-empirical potentials, in general, suffer from limited accuracy
and are not very flexible, because of their rather simple analytical form. This
cannot capture the properties of structures over a large phase space.

Machine learning (ML) based regression techniques~\cite{PhysRevLett.98.146401,PhysRevLett.104.136403,Behler_NN_review2017,Botu_ML_review_JPCC2017,Michele_PCCP2016,Marques_npj2019}
have recently emerged as a promising tool to construct  interatomic potentials. Their advantage is that they are entirely data-driven
and do not assume
any specific functional form. Most machine-learned force fields (MLFF) try to learn the potential
energy surface as well as its derivatives by finding a map from the local atomic environments onto local energies.
Typically, energies, forces, and stress tensors that are calculated by first-principles (FP) techniques are fitted.
Using the kernel ridge regression method, Zong~\emph{et al.} generated an interatomic potential
that successfully reproduced the phase diagram of Zr~\cite{Zong_ML_npj2018}
and uncovered the nucleation mechanism for the shock-induced hcp-bcc phase transformation in hcp-Zr~\cite{Zong_ML_PRB2020}.
Using the Gaussian approximation potential  model~\cite{PhysRevLett.104.136403,PhysRevB.87.184115},
Qian and Yang~\cite{Qian_ML_PRB2018} studied the temperature-induced phonon renormalization of bcc Zr
and clarified the origin of its instability at low temperature. However, for the hereto employed ML methods,
construction of suitable training structures is a fairly time-consuming trial and error process
based on intuition. The thus obtained training datasets are normally huge and might contain
unnecessary structures outside the phase space of interest. This can even reduce the accuracy
of the generated ML potential. Furthermore, the generated ML potential showed only
fair agreement with  phonon frequencies and elastic constants calculated using density functional theory (DFT).

To reduce human intervention, on-the-fly machine learning
schemes~\cite{PhysRevLett.114.096405,PhysRevLett.120.026102,PhysRevLett.122.225701} provide
an elegant solution. These generate the force fields automatically during FP MD
simulations while exploring potentially a large phase space. In particular,
Jinnouchi~\emph{et al.}~\cite{PhysRevLett.122.225701,PhysRevB.100.014105} suggested using the
predicted Bayesian error to judge whether FP calculations are required or not.
In this manner, usually more than 98\% of the FP calculations are bypassed during the training, significantly
enhancing the sampling of the configuration space and the efficiency of the force field generation~\cite{PhysRevLett.122.225701}.
This method has been successfully applied to the accurate and efficient prediction of
entropy-driven phase transitions of hybrid perovskites~\cite{PhysRevLett.122.225701},
melting points~\cite{PhysRevB.100.014105}, as well as chemical potentials of atoms and molecules~\cite{PhysRevB.101.060201}.

In this work, we attempt to revisit the hcp-bcc phase transition of Zr at ambient pressure
by using the on-the-fly MLFF method developed by
Jinnouchi~\emph{et al.}~\cite{PhysRevLett.122.225701,PhysRevB.100.014105}.
Almost without any human intervention, our generated MLFF successfully reproduces
the phonon dispersions of both the hcp and bcc phases at 0 K as well as the
first-order displacive nature of the phase transition manifested by an abrupt jump of the volume
and cooperative movement of atoms at the phase transition temperature. This confirms the Burgers mechanism~\cite{BURGERS1934561}.
The phase transition is further confirmed by the simulated x-ray powder diffraction.
Moreover, we demonstrate that using a singular value decomposition for the regression
overall improves the accuracy of the MLFF compared to the regularized Bayesian regression.

\section{Method}\label{sec:method}

For a comprehensive description of the on-the-fly MLFF generation implemented in the Vienna
\emph{Ab initio} Simulation Package (VASP), we refer to Ref.~\cite{PhysRevB.100.014105}.
A perspective article on this method can be found in Ref.~\cite{doi:10.1021/acs.jpclett.0c01061}.
Here, we just summarize the most important aspects of the underlying MLFF techniques.

As in many MLFF methods~\cite{PhysRevLett.98.146401,PhysRevLett.104.136403,Behler_NN_review2017,Botu_ML_review_JPCC2017, PhysRevB.87.184115, PhysRevB.90.024101,doi:10.1137/15M1054183,PhysRevB.97.184307,doi:10.1063/1.5020710, Michele_PCCP2016,Marques_npj2019},
the potential energy $U$ of a structure with $N_{a}$ atoms is approximated as a summation of local atomic potential energies $U_{i}$
\begin{align}\label{eq:local_approx}
U &= \sum\limits_{i=1}^{N_{a}} U_{i},
\end{align}
where $U_{i}$ is described as a functional of the two-body ($\rho_{i}^{(2)}$) and three-body ($\rho_{i}^{(3)}$) distribution functions,
\begin{align}\label{eq:functional}
U_{i} &= F \left[ \rho_{i}^{(2)}, \rho_{i}^{(3)} \right].
\end{align}
The two-body distribution function $\rho_{i}^{(2)}$ is defined as the probability to find an atom $j \left(j \neq i \right)$ at a distance $r$ from atom $i$~\cite{PhysRevB.100.014105,doi:10.1063/5.0009491}
\begin{align}\label{eq:distri_2}
\rho_{i}^{(2)}\left(r\right) &= \frac{1}{4\pi} \int \rho_{i}\left(r\hat{\mathbf{r}}\right) d\hat{\mathbf{r}},
\end{align}
where $\rho_{i}(\rr)$ ($\rr=r\hat{\mathbf{r}}$) is the three-dimensional atom distribution function around the atom $i$ defined as
\begin{align}
  \begin{split}
\rho_{i}\left(\mathbf{r}\right) &= \sum\limits_{j \neq i}^{N_{a}} \tilde  \rho_{ij}\left(\mathbf{r}\right), \\
\tilde  \rho_{ij}\left(\mathbf{r}\right) &= f_{\rm cut}\left(|\mathbf{r}_{j}-\mathbf{r}_{i}|\right) g\left(\mathbf{r}-(\mathbf{r}_{j}-\mathbf{r}_{i})\right).
\end{split}
\end{align}
Here, $\tilde  \rho_{ij}\left(\mathbf{r}\right)$ is the likelihood to find atom $j$ at position $\mathbf{r}$ relative to atom $i$,
$f_\textup{cut}$ is a cutoff function that smoothly eliminates the contribution from atoms outside a given cutoff radius $R_{\rm cut}$,
and $g$ is a smoothed $\delta$-function. The three-body distribution function
$\rho_{i}^{(3)}$ is defined as the probability to find an atom $j \left(j \neq i \right)$ at a distance $r$ from atom $i$ and another atom $k \left(k\neq i,j \right)$ at a distance $s$ from
atom $i$ spanning the angle $\angle{kij}=\theta$ between them. It is defined as~\cite{doi:10.1063/5.0009491}
\begin{equation}\label{eq:distri_3}
\begin{split}
\rho_{i}^{(3)}\left(r,s,\theta\right) &= \iint  d\hat{\mathbf{r}} d\hat{\mathbf{s}} \; \delta\left(\hat{\mathbf{r}}\cdot\hat{\mathbf{s}} - \mathrm{cos}\theta\right)
\sum\limits_{j \neq i}^{N_{a}} \sum\limits_{k\neq i, j}^{N_{a}} \tilde  \rho_{ik} \left( r\hat{\mathbf{r}} \right) \tilde \rho^*_{ij} \left( s\hat{\mathbf{s}} \right)  \\
 &= \iint d\hat{\mathbf{r}} d\hat{\mathbf{s}} \; \delta \left(\hat{\mathbf{r}}\cdot\hat{\mathbf{s}} - \mathrm{cos}\theta\right) \\
&  \quad  \quad \times \left[ \rho_{i}\left(r\hat{\mathbf{r}}\right) \rho_{i}^{*} \left(s\hat{\mathbf{s}}\right) - \sum\limits_{j\neq i}^{N_{a}} \tilde  \rho_{ij} \left( r\hat{\mathbf{r}} \right) \tilde  \rho^*_{ij} \left( s\hat{\mathbf{s}} \right) \right].
 \end{split}
\end{equation}
It should be noted that the definition of $\rho_{i}^{(3)}$ in Eq.~\eqref{eq:distri_3}   is free of two-body components
and the importance of the two- and three-body descriptors can thus be separately tuned.
To distinguish from the power spectrum~\cite{PhysRevB.87.184115}, we denote these descriptors as the separable descriptors.
For more discussions on the difference between the separable descriptors and the power spectrum, we refer to Ref.~\cite{doi:10.1063/5.0009491}.

In practice, $\rho_{i}^{(2)}$ and $\rho_{i}^{(3)}$ are discretized in a suitable basis
and represented by a descriptor vector $\mathbf{x}_{i}$ collecting all two- and three-body coefficients~\cite{doi:10.1063/5.0009491}.
Therefore, the functional $F$ in Eq.~\eqref{eq:functional} becomes a function of $\mathbf{x}_{i}$~\cite{doi:10.1063/5.0009491}
\begin{equation}
U_{i} = F \left[ \rho_{i}^{(2)}, \rho_{i}^{(3)} \right] \rightarrow  F(\mathbf{x}_{i} ).
\end{equation}
For the functional form of $F$, a kernel based approach is used~\cite{PhysRevB.87.184115}.
Specifically, using the algorithm of data selection and sparsification~\cite{PhysRevB.100.014105},
$N_B$ atoms are chosen from a set of reference structures generated by FP MD simulations
and the atomic distributions surrounding the selected atoms are mapped onto the descriptors $\mathbf{x}_{i_B}$.
The function $F$ is then approximated by the linear equation of coefficients $w_{i_B}$
\begin{align}\label{eq_kernel}
F(\mathbf{x}_{i} ) &=\sum\limits_{i_{B}=1}^{N_{B}} w_{i_{B}} K\left(\mathbf{x}_{i},\mathbf{x}_{i_{B}} \right),
\end{align}
where the kernel function $K\left(\mathbf{x}_{i},\mathbf{x}_{i_{B}} \right)$ is a nonlinear function that
is supposed to quantify the degree of similarity between a local configuration
$\mathbf{x}_{i}$ of interest and the reference configuration $\mathbf{x}_{i_{B}}$.
Here, a polynomial function $K\left(\mathbf{x}_{i},\mathbf{x}_{i_{B}} \right)= \left( \mathbf{\hat{x}}_{i} \cdot \mathbf{\hat{x}}_{i_{B}} \right)^{\zeta}$
is used~\cite{PhysRevB.87.184115, doi:10.1063/5.0009491}.
This introduces nonlinear mixing of purely two- and three-body descriptors, which was found to be important
for an accurate and efficient description of the potential energy surfaces~\cite{doi:10.1063/5.0009491}.

From Eq.~\eqref{eq_kernel}, the total energy, forces and stress tensors of any structure
can be obtained as linear equations of the coefficients $w_{i_B}$.
In a matrix-vector representation, it can be expressed  as
\begin{align}\label{eq_linear_equation}
\mathbf{y}^{\alpha} & = \bm{\phi}^{\alpha}\mathbf{w},
\end{align}
where $\mathbf{y}^{\alpha}$ is a vector collecting the FP energy, forces, and stress tensors for the given structure $\alpha$ of $N^{\alpha}_{a}$ atoms,
in total, $m^\alpha=1+3N^{\alpha}_{a}+6$ components.
$\bm{\phi}^{\alpha}$ is a $m^\alpha \times N_B$ matrix. The first line of the matrix $\bm{\phi}^{\alpha}$ is comprised of
$\sum_i^{N^{\alpha}_{a}} K\left(\mathbf{x}^{\alpha}_{i},\mathbf{x}_{i_\mathrm{B}}\right)/ N^{\alpha}_{a}$,
the subsequent $3N^{\alpha}_{a}$ lines consist of the derivatives of the kernel with respect to the atomic coordinates,
and the final six lines consist of the derivatives of the kernel with respect to the unit cell coordinates~\cite{PhysRevB.100.014105}.
$\mathbf{w}$ is a vector collecting all coefficients $\{ w_{i_{B}} | i_{B} =1,...,N_{B} \}$.
The generalized linear equation containing all reference structures is given by
\begin{align}\label{eq_linear_equation_all}
\mathbf{y}=\mathbf{\Phi}\mathbf{w}.
\end{align}
Here, ${\mathbf{y}}$ is a super vector collecting all FP energies, forces, and stress tensors
$\{\mathbf{y}^{\alpha}|\alpha=1,...,N_\mathrm{st}\}$ for all reference structures
and similarly, $\mathbf{\Phi}$ is the design matrix
comprised of matrices $\bm{\phi}^{\alpha}$ for all reference structures~\cite{PhysRevB.100.014105}.
Based on Bayesian linear regression (BLR),  the optimal coefficients $\mathbf{\bar{w}}$ are determined as~\cite{PhysRevB.100.014105,book_Bishop}
\begin{equation}\label{eq:coefficients}
\mathbf{\bar{w}} =
 \Big( \mathbf{\Phi}^{\mathrm{T}}\mathbf{\Phi}+ {\sigma_{\mathrm{v}}^{2}}/{\sigma_{\mathrm{w}}^{2}} \mathbf{I} \Big)^{-1}
 \mathbf{\Phi}^{\mathrm{T}} \mathbf{y},
\end{equation}
where $\sigma_{\mathrm{v}}^{2}$ is the variance of the uncertainty caused by noise in the training datasets,
and $\sigma_{\mathrm{w}}^{2}$ is the variance of the prior distribution~\cite{PhysRevB.100.014105}.
The parameters $\sigma_{\mathrm{v}}^{2}$ and $\sigma_{\mathrm{w}}^{2}$ are obtained by maximizing the evidence function~\cite{PhysRevB.100.014105}.

Having obtained the optimal coefficients $\mathbf{\bar{w}}$, the energy, forces, and stress tensors for any given structure $\alpha$
can be predicted by $\mathbf{y}^{\alpha} = \bm{\phi}^{\alpha}\mathbf{\bar{w}}$, and the uncertainty in the prediction
is estimated as the variance of the posterior distribution~\cite{doi:10.1021/acs.jpclett.0c01061}
 \begin{align}\label{eq_sigma}
\bm{\sigma}^2=\sigma_{\mathrm{v}}^{2}\mathbf{I}+\sigma_{\mathrm{v}}^{2} \bm{\phi}^{\alpha}
\Big(\mathbf{\Phi}^{\mathrm{T}}\mathbf{\Phi}+ {\sigma_{\mathrm{v}}^{2}}/{\sigma_{\mathrm{w}}^{2}} \mathbf{I} \Big)^{-1} [\bm{\phi}^{\alpha}]^{\mathrm{T}}.
\end{align}
It is found that the square root of the second term in Eq.~\eqref{eq_sigma} resembles the real error remarkably well~\cite{PhysRevB.100.014105} and thus
provides a reliable measure of the uncertainty. This is the heart of the on-the-fly MLFF algorithm.
Armed with a reliable error prediction, the machine can decide whether new structures
are out of the training dataset or not by using state-of-the-art query strategies~\cite{PhysRevB.100.014105}.
Only if the machine finds the need to update the training dataset with the new structures, then FP calculations are carried out.
Otherwise, the predicted energy, forces, and stress tensors by the yet available MLFF
are used to update the atomic positions and velocities. In this manner, most of the FP calculations are bypassed during training runs
and simulations are in general accelerated by several orders of magnitude while retaining almost FP accuracy~\cite{PhysRevB.100.014105,doi:10.1021/acs.jpclett.0c01061}. A final note is in place here: we generally distinguish between training runs and the final application of the MLFF.
In the first case, the force field is continuously updated and the total energy is not a constant of motion, whereas in the latter
this is the case.

Furthermore, we notice that in Eq.~\eqref{eq:coefficients}, disregarding regularization,
essentially an inversion of a squared matrix $\mathbf{\Phi}^{\mathrm{T}}\mathbf{\Phi}$ is performed
\begin{equation}\label{eq:normal_eq}
\mathbf{\bar{w}} =  \Big( \mathbf{\Phi}^{\mathrm{T}}\mathbf{\Phi} \Big)^{-1}  \mathbf{\Phi}^{\mathrm{T}} \mathbf{y}.
\end{equation}
Similar procedures (inversion of a squared matrix) are adopted by Cs\'{a}nyi and coworkers~\cite{PhysRevB.90.104108},
although a different regularization is used.
We find that the condition number of the squared matrix $\mathbf{\Phi}^{\mathrm{T}}\mathbf{\Phi}$
often approaches $1/\epsilon$, where $\epsilon$ is the machine precision (for double precision arithmetic $\epsilon$ is roughly $10^{-16}$).
Squaring the matrix $\mathbf{\Phi}$, i.e., calculating $\mathbf{\Phi}^{\mathrm{T}}\mathbf{\Phi}$ means that the condition number of the
matrix $\mathbf{\Phi}$ is also squared. If the condition number of the squared matrix exceeds $1/\epsilon$,
information is irrevocably lost from the original problem. The standard means
to avoid squaring the problem is to replace the solution of the normal equation~\eqref{eq:normal_eq}
by the $QR$ decomposition $\mathbf{\Phi} = \mathbf{Q}\mathbf{R}$ and to obtain $\mathbf{\bar{w}}$ by backwards substitution
$\mathbf{R} \mathbf{\bar{w}} = \mathbf{Q}^{\mathrm{T}}\mathbf{y}$.
It is well known that $QR$ algorithms significantly improve the stability of the solution of a least square problem.
A slightly more expensive and equally controlled solution is to calculate the pseudoinverse of $\mathbf{\Phi}$ using a singular value decomposition (SVD)
\begin{align}
\mathbf{\Phi} &= \mathbf{U} \mathbf{\Sigma} \mathbf{V}^\mathrm{T},  \\
\mathbf{\Phi}^{-1} & =\mathbf{V} \mathbf{\Sigma}^{-1}\mathbf{U}^\mathrm{T}, \\
 \mathbf{\bar{w}} &=\mathbf{\Phi}^{-1}\mathbf{y}.\label{eq:coefficients_SVD}
\end{align}
This can be calculated by calling scaLAPACK routines~\cite{ScaLAPACK}.
The key question is whether this allows us to salvage the additional information from $\mathbf{\Phi}$ that is lost by squaring the problem and solving
the regularized normal equation. Inspection of the eigenvalue spectrum of $\mathbf{\Phi}$ for the
present case shows that the condition number of $\mathbf{\Phi}$ is roughly $4\times10^{10}$. This confirms
that some information is lost due to extinction and finite precision in the matrix $\mathbf{\Phi}^{\mathrm{T}}\mathbf{\Phi}$, which
would formally have a condition number of $\approx$$10^{21}$.
As we show below, we are indeed able to recover some additional information and thus improve
the accuracy by calculating the pseudoinverse of the design matrix instead of solving the
regularized normal equation. In the present case the advantages are, however, small.
Since calculating the pseudoinverse takes little extra time, we feel that
this step should be performed regardless of the admittedly small gains in accuracy.
We do this only once, after the on-the-fly training has finished.
We note that the condition number of the design matrix $\mathbf{\Phi}$ is rarely
reported in literature for MLFFs. It would be interesting to know whether other implementations
observe similar issues. Specifically, we expect that a combination of radial and angular
descriptors or use of the power spectrum generally leads to a fairly ill-conditioned problem.

Finally, we stress that regularization in a manner strictly compatible to Eq.~\eqref{eq:coefficients}
can be easily recovered by using the Tikhonov regularization~\cite{book_Tikhonov} if needed.
However, contrary to common belief, we find that due to the inclusion of equations for the forces and sparsification of the local environments,
our system of equations is in general overdetermined and therefore regularization is not strictly required.
To give an example, in the present case, the final force field is trained using 935 structures of 48 atoms, each yielding one energy
equation, six equations for the stress tensor, and $48\times3$ equations for the forces. Due to sparcification
only 1013 fitting coefficients need to be determined (see Sec. \ref{sec:training_details}).
This means that the number of equations is about 140 times larger
than the number of unknowns.
Finally, we note that we use the evidence approximation to determine $\sigma_{\mathrm{v}}^2$ and $\sigma_{\mathrm{w}}^2$.
We find that the quotient $(\sigma_{\mathrm{v}}^2/\sigma_{\mathrm{w}}^2)/\lambda_{\rm max}$
($\lambda_{\rm max}$  being the maximum eigenvalue of the squared matrix $\mathbf{\Phi}^{\mathrm{T}}\mathbf{\Phi}$)
approaches machine precision in the present case. This also confirms that the system of equations is overdetermined and that regularization is not required.

\section{Computational Details}\label{sec:details}

\subsection{First-principles calculations}\label{sec:DFT_details}

All FP calculations were performed using VASP~\cite{PhysRevB.47.558, PhysRevB.54.11169}.
The generalized gradient approximation of Perdew-Burke-Ernzerhof (PBE)~\cite{PhysRevLett.77.3865} was used for the exchange-correlation functional.
A plane wave cutoff of 500 eV and a $\Gamma$-centered $k$-point grid with a spacing of 0.16 $\AA^{-1}$ between $k$ points were employed,
which ensure that the total energy is converged to better than 1 meV/atom. The Gaussian smearing method with a smearing width
of 0.05 eV was used to handle fractional occupancies of orbitals in the Zr metal. The electronic optimization was performed until the total energy
difference between two iterations was less than 10$^{-6}$ eV.

\subsection{MLFF training}\label{sec:training_details}

Our MLFFs were trained on-the-fly during MD simulations using a Langevin thermostat~\cite{book_Allen_Tildesley} at ambient pressure
with a time step of 1.5 fs. The separable descriptors~\cite{doi:10.1063/5.0009491} were used.
The cutoff radius for the three-body descriptor and the width of the Gaussian functions used for broadening the atomic
distributions of the three-body descriptor were set to 6 $\AA$ and 0.4 $\AA$, respectively. The number of radial basis
functions and maximum three-body momentum quantum number of spherical harmonics used to expand the atomic distribution
for the three-body descriptor were set to 15 and 4, respectively. The parameters for the two-body descriptor were
the same as those for the three-body descriptor.

The training was performed on a 48-atom orthorhombic cell using the following strategy.
($i$) We first trained the force field by a heating run from 0 to 1600 K using 20 000 MD steps starting from the DFT relaxed hcp structure.
($ii$) Then, we continued training the bcc phase by a MD simulation with an isothermal-isobaric (NPT) ensemble at $T$=1600 K using 10 000 MD steps.
($iii$) Using the equilibrium bcc structure at $T$=1600 K obtained from the previous step, the force field was further trained  by a cooling run from 1600 to 0 K using 20 000 MD steps.
($iv$) Since the bcc Zr is strongly anharmonic and dynamically stable only at high
temperatures~\cite{Stassis_phonon_PRL1978,Qian_ML_PRB2018,Ye_phonon_cal_PRL1987,Heiming_phonon_Exp1991,Souvatzis_PRL2008},
to include the ideal 0 K bcc structure in the training dataset, an additional heating run from 0 to 300 K using 10 000 MD steps
was performed starting from the DFT relaxed bcc structure. Indeed, we observed that
the bcc phase is unstable at low temperature and transformed into the more stable hcp structure just after 300 MD steps.
It should be stressed here that our on-the-fly MLFF training is rather efficient.
Eventually, only 935 FP calculations were performed out of 60 000 MD steps, i.e., nearly 98.4\%  of the FP calculations were bypassed.
From these 935 reference structures, 1013 local configurations are selected as the basis sets. In the last step, the SVD  [Eq.~\eqref{eq:coefficients_SVD}]
was used to redetermine the coefficients using the same design matrix as obtained from the BLR.
In the following, we denote the MLFFs obtained by using BLR and SVD for the regression as MLFF-BLR and MLFF-SVD, respectively.

\begin{table}
\caption {The training and validation RMSE in energies (meV/atom), forces (eV/$\AA$) and stress tensors (kbar)
calculated by MLFF-BLR and MLFF-SVD for three $\omega_E$. Note that in this work $\omega_E$=10 is used unless otherwise explicitly stated.}
\begin{ruledtabular}
\begin{tabular}{llccccc}
& & \multicolumn{2}{c}{Training errors} & & \multicolumn{2}{c}{Validation errors} \\
 \cline{3-4}  \cline{6-7}
        &         &  BLR  &  SVD   & &      BLR  &   SVD  \\
                  \hline
\multirow{3}{1.5cm}{$\omega_E=1$ }
& Energy &  3.69 &  3.22  & &      2.87 &   2.70 \\
& Force   &  0.08 &  0.07  & &      0.10 &   0.09 \\
& Stress    &  1.16 &  1.04  & &      1.16 &   1.12 \\
\hline
\multirow{3}{1.5cm}{$\omega_E=10$ }
& Energy &  2.33 &  1.74  & &      2.17 &   1.96 \\
& Force   &  0.08 &  0.07  & &      0.10 &   0.09 \\
& Stress    &  1.40 &  1.05  & &      1.34 &   1.11 \\
\hline
\multirow{3}{1.5cm}{$\omega_E=100$ }
& Energy &  1.65 &  0.47  & &      2.87 &   2.36  \\
& Force   &  0.09 &  0.08  & &      0.11 &   0.10 \\
& Stress    &  1.89 &  1.27  & &      1.98 &   1.29 \\
\end{tabular}
\end{ruledtabular}
\label{tab:error}
\end{table}

Furthermore, we note that for any regression method it is possible to increase the weight of some equations,
though this reduces the ``relevance" and in turn the accuracy of the other equations.
Presently our machine learning code first reweights all equations such that the standard deviations
in the energy per atom, forces and stress tensors equal one. To give an example, if the standard
deviation in the energy per atom is 100 meV, all energy equations are scaled by 1/100 meV$^{-1}$. Likewise,
if the standard deviation for the forces is 0.5 eV/\AA, all force equations are scaled by 2 (eV/\AA)$^{-1}$.

After this scaling has been performed, we found that it is expedient to increase the relative weight of the energy equations ($\omega_E$) by a factor of 10
with respect to the equations for the forces and stress tensors in the linear regression. This decreased the root-mean-squared errors (RMSE) in the energies by
almost 1.4 meV/atom for the training dataset, while the errors in the forces and stress
tensors did not increase significantly (see Table~\ref{tab:error}). One motivation for increasing $\omega_E$ is
that for each structure with $N_a$ atoms, there is only one equation for the energy,
but 3$N_a$ and 6 equations for the forces and stress tensors, respectively. Likewise,
we found that increasing the relative weight of the stress tensor equations ($\omega_S$) by a factor of 5
improves the accuracy of the elastic constants, although it slightly worsens phonon dispersion relations (see Sec.~\ref{sec:results}).

\subsection{MLFF validation}\label{sec:validation_details}

\begin{figure}
\begin{center}
\includegraphics[width=0.48\textwidth,clip]{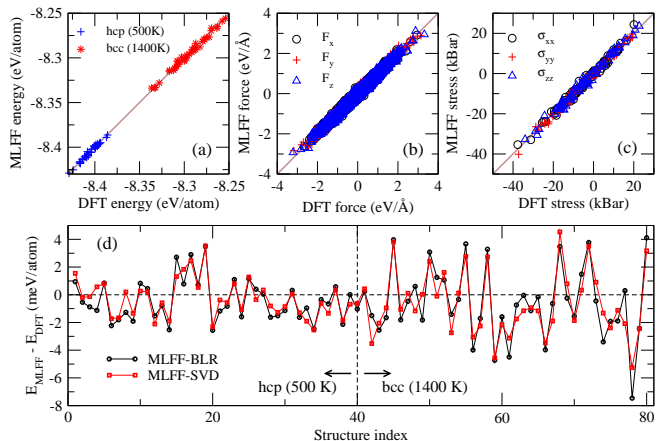}
\end{center}
 \caption{MLFF-SVD  vs. DFT in terms of (a) energies, (b) forces, and (c) diagonal components of the stress tensors
 for the test datasets. (d) The MLFFs and DFT predicted energy difference for each structure in the test datasets.
}
\label{fig:Error_analysis}
\end{figure}

The generated MLFFs have been validated on a test dataset containing 40 hcp structures of 64 atoms
at $T$=500 K and another 40 bcc structures of 64 atoms at $T$=1400 K. These structures were generated
using MD simulations with an NPT ensemble at $T$=500 and 1400 K employing the obtained MLFFs.
Table~\ref{tab:error} shows both the training and validation errors in energies, forces, and stress tensors
calculated by MLFF-BLR and MLFF-SVD. Clearly, results using SVD are generally improved compared
to the results using BLR, both for the test and training dataset. Although the improvement seems
to be modest, we will see below that physical observables are also better described using the SVD.
Concerning the relative weight of the energy equations, we note that using SVD the error in the energy
in the training dataset decreases significantly, reaching sub meV precision (0.47 meV/atom), if
the energy equations are reweighted by a factor of 100.
Unfortunately, the errors in the test dataset increase, if $\omega_E$
is increased beyond a value of 10. This indicates that by strongly weighting the energy equations,
the unregularized SVD tends to overfit the energies, and overall the best results on the test dataset are
obtained by reweighting the energy equations by a factor of 10 and using SVD.

As an illustration, results
on the energies, forces and diagonal components of stress tensors predicted by MLFF-SVD and density functional theory (DFT) for the
test dataset are presented in Figs.~\ref{fig:Error_analysis}(a)$-$\ref{fig:Error_analysis}(c), respectively, showing very good agreement.
In addition, the MLFFs and DFT predicted energy difference for each structure in the test datasets is shown in Fig.~\ref{fig:Error_analysis}(d).
Compared to the hcp structures, the bcc ones exhibit larger errors due to the stronger thermal fluctuations at high temperature.
We note that our generated MLFF-BLR is already very accurate with training and
validation errors of 2.33 and 2.17 meV/atom in the energy, respectively.
Due to the improved condition number, MLFF-SVD further improves upon
MLFF-BLR by reducing the overall errors in energies, forces and stress tensors (see Table~\ref{tab:error}).
These improvements are particularly relevant  for the application to the prediction of defects energetics
where supercells need to be used and errors in the range of 1 meV/atom will cause errors of
the order of 100 meV for defects. In addition, as compared to MLFF-BLR,
MLFF-SVD improves the phonon dispersions towards DFT results due to its improved forces, as will be discussed later on.

\begin{table}
\caption {Lattice parameters of hcp and bcc Zr as well as their energy difference at 0 K predicted by DFT and MLFFs using
BLR and SVD for the regression. Note that the experimental data for hcp Zr~\cite{Kolesnikov_1994} and
bcc Zr~\cite{doi:10.1143/JPSJ.36.1349} were measured at room temperature and low temperature ($<$ 7 K), respectively.
}
\begin{ruledtabular}
\begin{tabular}{lcccc}
& DFT &  BLR & SVD  & Expt. \\
\hline
hcp Zr & &  & &   \\
 $a=b$ (\AA) & 3.235  & 3.234 & 3.235 & 3.233~\cite{Kolesnikov_1994} \\
 $c$ (\AA) & 5.167 &5.169 &  5.166 & 5.147~\cite{Kolesnikov_1994} \\
 \hline
bcc Zr & & & & \\
 $a$ (\AA)  & 3.574 & 3.574 & 3.573 & 3.551~\cite{doi:10.1143/JPSJ.36.1349} \\
 \hline
$E$(bcc)$-E$(hcp) (eV/atom) &0.084 &  0.081 & 0.082 &  ---
\end{tabular}
\end{ruledtabular}
\label{tab:latt-par}
\end{table}

We notice that our force field is more accurate than the one obtained by Zong \emph{et al.}~\cite{Zong_ML_npj2018},
which exhibited much larger training mean absolute errors of 5.8 and 6.7 meV/atom in the energy for hcp and bcc Zr, respectively.
This might be related to the fairly simplified ML model used in Ref.~\cite{Zong_ML_npj2018} as well
as a rather extensive training dataset containing multiphase structures.
Surprisingly, the force field generated by Qian and Yang~\cite{Qian_ML_PRB2018} shows
rather small validation RMSE of 0.2 meV/atom for the hcp phase and 0.3 meV/atom for the bcc phase~\cite{Qian_ML_PRB2018}.
In our experience, a precision of sub meV/atom can only be attained
if fairly small displacements and low temperature structures are used. Indeed,
the training structures considered in Ref.~\cite{Qian_ML_PRB2018} correspond to small displacements
of the groundstate hcp and bcc structure as well as finite temperature training data at 100, 300, and 1200~K,
and validation was done for configurations selected from MD simulations at 300~K.

\section{Results}\label{sec:results}

We start by showing the lattice parameters of hcp and bcc Zr at 0 K
as well as their energy difference predicted by DFT and MLFFs.
As seen in Table~\ref{tab:latt-par}, almost perfect agreement is observed
between DFT and MLFFs for both BLR and SVD. The slightly larger lattice parameters predicted by theory as compared to experiment
originate from the tendency of PBE to overestimate lattice constants. For the energy difference between bcc and hcp Zr,
both MLFF-BLR and MLFF-SVD slightly underestimate the DFT value with MLFF-SVD being more accurate (see also Table~\ref{tab:error}).

\begin{figure}
\begin{center}
\includegraphics[width=0.4\textwidth,trim = {0 0.25cm 5cm  0.0cm}, clip]{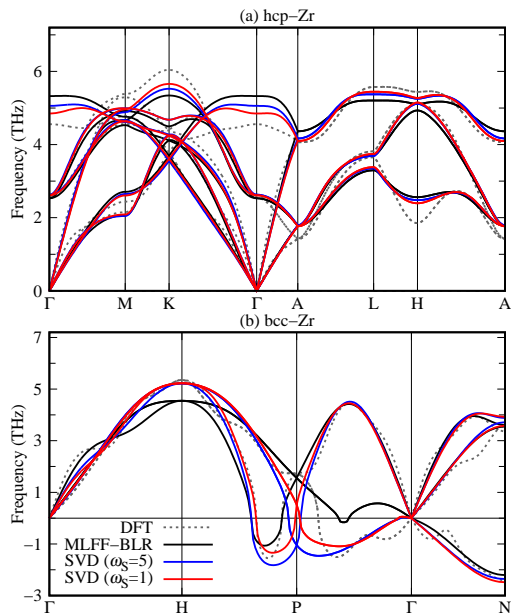}
\end{center}
 \caption{Phonon dispersion relation of (a) hcp and (b) bcc Zr at 0 K predicted by DFT (grey dotted lines) and MLFFs (full lines)
 using BLR (black) and SVD (red for $\omega_S$=1 and blue for $\omega_S$=5) for the regression.
 180-atom and 343-atom supercells have been used for hcp and bcc phases, respectively.
}
\label{fig:phonons}
\end{figure}

Figure~\ref{fig:phonons} presents the phonon dispersions of hcp and bcc Zr at 0 K calculated by DFT and MLFFs.
Consistent with previous FP calculations~\cite{Chen_phonon_cal_PRB1985,Souvatzis_PRL2008,Zong_ML_npj2018,Qian_ML_PRB2018},
at 0 K hcp Zr is dynamically stable, whereas bcc Zr is dynamically unstable
due to the double-well shape of the potential energy surface~\cite{Qian_ML_PRB2018}.
As compared to DFT, MLFF-BLR describes the acoustic phonons of hcp Zr very well.
Although a slightly larger deviation exists for the optical phonons,
it seems that difficulties in accurately describing optical phonons
are quite general for machine learned interatomic potentials~\cite{Zong_ML_npj2018,Qian_ML_PRB2018}.
For instance, our results are comparable with those predicted by Qian and Yang~\cite{Qian_ML_PRB2018}, but
are better than those predicted by Zong \emph{et al.}~\cite{Zong_ML_npj2018}. The latter
show a very large discrepancy of nearly 2 THz for the optical phonons at the Brillouin-zone center~\cite{Zong_ML_npj2018}.
The possible reasons have been discussed in Sec.~\ref{sec:validation_details}. Here, we want to emphasize that
in contrast to Ref.~\cite{Qian_ML_PRB2018} where the force field was purposely  trained
to model phonons by using perturbed supercells with strains and displacements, in the present work,
the necessary information on the  force constants were automatically
captured during the on-the-fly MLFF training, and our MLFF predicted phonon dispersions came out to
be in good agreement with the DFT results.
In addition, we observe that the average optical phonon frequencies predicted by our MLFFs are quite accurate,
which implies that free energy differences are likely to be described accurately.
For the bcc phase, the MLFF-BLR is able to capture the soft zone-boundary $N$-point phonon of
the $T_1$ branch which is involved in the $\beta$-$\alpha$ phase transition~\cite{BURGERS1934561,Petry_Exp1991}
and the soft phonon mode in the $H$-$P$ direction that is responsible for
the $\beta$-$\omega$ phase transition~\cite{Stassis_phonon_PRL1978,Petry_Exp1991,Sikka_review1982},
but struggles to obtain accurate results along $P$-$\Gamma$. However, these soft phonon modes are extremely difficult
to obtain accurately even by DFT, with the DFT results being strongly dependent on the system size.
This means that training on a 48-atom cell is likely to be inadequate to describe all phonon instabilities in bcc Zr.
As compared to MLFF-BLR, MLFF-SVD overall improves the phonon dispersions towards the DFT results
for both hcp and bcc Zr, in particular for the optical phonon modes for both phases
and the soft phonon modes along $P$-$\Gamma$  for bcc Zr. This is not unexpected, since
MLFF-SVD reduces errors in forces as compared to MLFF-BLR (see Table~\ref{tab:error}).

\begin{figure}
\begin{center}
\includegraphics[width=0.4\textwidth,clip]{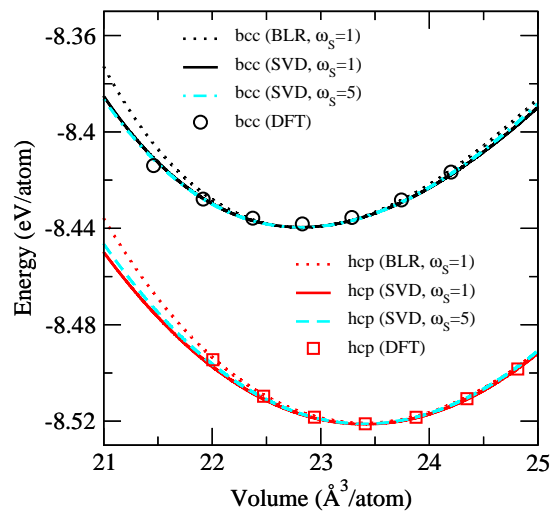}
\end{center}
 \caption{Energies of hcp and bcc Zr at 0 K as a function of volume predicted by DFT and MLFFs.
Curves are fitted by using the Vinet universal equation of state~\cite{Vinet_1989}.
}
\label{fig:E_V_compare}
\end{figure}

\begin{table}
\caption {Ion-relaxed elastic coefficients $C_{ij}$ and bulk moduli (in GPa) of hcp and bcc Zr at 0 K predicted by DFT and MLFFs.
For the MLFFs using SVD, results using two relative weights of the stress tensor equations ($\omega_S$=1 and 5) are shown.
Values in the parentheses represent the ion-clamped elastic coefficients.
The experimental data of hcp Zr~\cite{Fisher_Exp1964} and  bcc Zr~\cite{Heiming_phonon_Exp1991}
shown in this table were measured at 4 and 1189 K, respectively.
}
\begin{ruledtabular}
\begin{tabular}{lccccc}
& DFT &  BLR & SVD  & SVD & Expt. \\
&  &  ($\omega_S$=1) & ($\omega_S$=1)  & ($\omega_S$=5) & \\
\hline
hcp Zr & &  &   \\
$C_{11}$  & 143.4  & 133.9 & 132.0 & 140.7 &155.4   \\
                    & (155.3)  & (150.3) & (142.9) & (149.6) & --   \\
$C_{12}$  & 64.9  & 78.0 & 61.8 &63.2 & 67.2   \\
                   & (52.9)  & (61.5) & (50.8) & (54.3) & --   \\
$C_{13}$  & 65.4  & 68.6 &56.7 & 62.8 &  64.6   \\
$C_{33}$  & 169.3  & 169.4 &158.9 & 158.0 & 172.5   \\
$C_{44}$  & 24.4  & 26.7& 24.5 & 24.0 &36.3   \\
$B$  & 93.31 & 95.55 & 89.0 & 91.8 &97.5   \\
 \hline
bcc Zr & &  & \\
$C_{11}$  & 73.3  & 109.0 & 72.3 & 76.7 & 104\\
$C_{12}$  & 95.3  & 117.8 & 105.4 & 108.1 &93 \\
$C_{44}$  & 28.8  & 39.5 & 36.5 & 32.7 &  38 \\
$B$  & 88.80 & 111.67& 97.1 & 97.3 &--- \\
\end{tabular}
\end{ruledtabular}
\label{tab:elastic}
\end{table}

Another important quantity for the prediction of phase transition are the elastic properties,
which are typically hard to accurately predict~\cite{Zong_ML_npj2018,Qian_ML_PRB2018,WIMMER2020152055}.
Although our MLFFs were trained during a heating/cooling MD simulation at a constant zero pressure only
(the focus of the present study is on the temperature-induced hcp-bcc phase transition at ambient pressure),
it turns out that the fluctuations of the volumes in the MD simulation allow us to sample slightly strained structures
and therefore our MLFFs are capable to describe elastic properties quite well.
Indeed, Fig.~\ref{fig:E_V_compare} shows the volume dependence of the energies of hcp and bcc Zr at 0 K predicted by DFT and MLFFs.
One observes that the DFT calculated energy vs. volume curve is well reproduced by our MLFFs.
Obvious deviations are discernible only for small volumes away from the equilibrium volume.
This is expected, because no external pressure is applied during training.
The better agreement between DFT and MLFFs for the larger volumes apparently benefits from
the thermal expansion during heating. As compared to the results in Ref.~\cite{Zong_ML_npj2018},
our MLFFs predicted energy vs. volume curves are, again, in better agreement with the DFT data.
Table~\ref{tab:elastic} summarizes the predicted elastic coefficients and bulk moduli.
One can see that our MLFFs work well for the elastic properties of hcp Zr, showing reasonably good agreement with DFT.
However, the description of the elastic properties for bcc Zr by our MLFFs is not so satisfactory. The largest discrepancy is
found for $C_{44}$. This is because at 0 K, the bcc phase is unstable both dynamically [see Fig.~\ref{fig:phonons}(b)]
and mechanically [the Born elastic stability criterion ($C_{11}-C_{12}>0$)~\cite{book_Born_Huang_1954} is disobeyed],
and therefore, only few reference structures corresponding to the unstable ideal bcc phase
are collected during our on-the-fly training.
Concerning the comparison between MLFF-BLR and MLFF-SVD,
we found that both MLFFs are comparably good in predicting the elastic properties of hcp Zr,
whereas the MLFF-SVD dramatically improves over the MLFF-BLR for bcc Zr. In addition, by increasing $\omega_S$
by a factor of 5, the overall elastic properties are further improved,
but this slightly worsens the phonon dispersion relations (see Fig.~\ref{fig:phonons}).
This is expected, because increasing $\omega_S$ yields more accurate stress tensors,
while slightly increasing the errors in energies and forces.

\begin{figure}
\begin{center}
\includegraphics[width=0.45\textwidth,trim = {0 0.1cm 5cm  0.5cm}, clip]{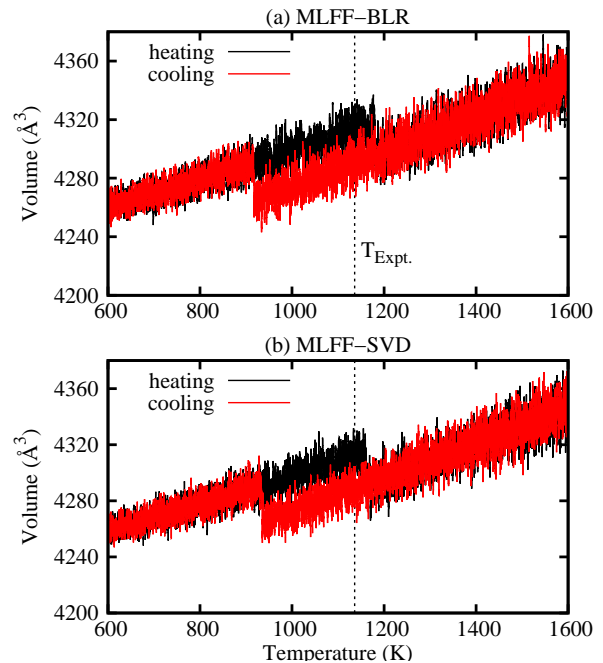}
\end{center}
 \caption{Evolution of the volumes of 180-atom orthorhombic supercells with respect to
 temperature during the heating (black) and cooling (red) MD simulations using (a) MLFF-BLR
 and (b) MLFF-SVD. The dashed lines represent the experimentally measured hcp-bcc phase transition temperature $T_{\rm Expt.}=1136$ K~\cite{Vogel1931_T1136K}.
}
\label{fig:Volume_vs_T}
\end{figure}

Finally, we turn to the hcp-bcc phase transition. To avoid large volume fluctuations appearing in small supercells,
a reasonably large orthorhombic supercell with 180 Zr atoms is used to simulate the phase transition.
Figure~\ref{fig:Volume_vs_T} shows the evolution of the volume with respect to
the temperature during the heating and cooling MD simulations
predicted by MLFF-BLR and MLFF-SVD.
For each MD simulation, 2 million MD steps (corresponding to a heating/cooling rate 0.33 K/ps) were used.
First, one can observe that both MLFFs successfully reproduce the hcp-bcc phase transition,
a typical first-order phase transition manifested by an abrupt jump in the volume
at $T_c$. Second, the predicted phase transition between
hcp and bcc phases is reversible via heating or cooling, but a fairly large hysteresis is observed, i.e.,
heating and cooling runs yield different $T_c$. This is not unexpected for a first-order phase transition
and similar to experimentally observed superheating and supercooling.
Third, if we average over the upper and lower transition temperatures,
both MLFFs predict a $T_c$ that is in reasonable agreement with the experimental value.
However, as compared to the phonon dispersion relations, no improvement for the
prediction of $T_c$ by SVD is obvious. We will explain this observation below.

We note that a quantitative comparison of $T_c$ between experiment and theory as obtained from direct heating and cooling should be done cautiously.
For small systems, the transition temperatures might well be wrong by 100 K due to errors introduced by finite size effects or hysteresis.
To mitigate this problem, we performed each heating or cooling run ten times to obtain a reasonable statistics for estimating $T_c$,
and we obtained a mean value of 1040 K with a standard deviation of 30 K for MLFF-SVD.
Upon heating the lowest temperature at which the transition to the bcc structure occurred was 1107 K, while
upon cooling the highest temperature at which the transition to the hcp structure occurred was 982 K.
The mean (1045 K) is in excellent agreement with the above value.
To refine the transition temperature further,
we lowered the heating and cooling rate by a factor of 4 (0.08 K/ps)
and performed four more cooling runs yielding transition temperatures of 1019-1047 K,
as well as four more heating runs yielding transition temperatures of 1046-1093 K. These values clearly confirm
that the transition temperature for a system size of 180 atoms is about 1045 K with an estimated error bar well below 10 K.
Such a small error bar would be very hard to achieve using, for instance, thermodynamic integration and free energy methods.

Finally, we have explored how accurate the  force fields, MLFF-BLR and MLFF-SVD, are compared to the
reference PBE calculation. The previous assessments on the ideal hcp and bcc structures are not necessarily very accurate,
since bcc Zr at 0 K is dynamically unstable, and finite temperature displacements are obviously not considered.
To assess the accuracy of the MLFF for predictions of the transition temperature, we estimate the free energy difference $F_\mathrm{FP}-F_\mathrm{MLFF}$
between FP and MLFF calculations through thermodynamic perturbation theory (TPT)
in the second-order cumulant expansion~\cite{doi:10.1063/1.1740409,PhysRevLett.121.195701}
\begin{equation}\label{eq:TPT2}
\begin{split}
F_\mathrm{FP}-F_\mathrm{MLFF}
&= -\frac {1} {k_{B} T} \mathrm{ln} \langle \mathrm{exp} \left( -\frac {U_{\mathrm{FP}} - U_{\mathrm{MLFF}}} {k_{\mathrm{B}} T} \right) \rangle \\
& \approx \langle \Delta U \rangle - \frac {1} {2k_{B} T} \langle (\Delta U - \langle \Delta U \rangle )^2 \rangle,
\end{split}
\end{equation}
where $\Delta U = U_{\mathrm{FP}} - U_{\mathrm{MLFF}}$ is the potential energy difference between FP and MLFF calculations.
Without loss of generality, 40 structures close to $T$=1040 K from the heating and cooling MD runs using MLFF-SVD
were selected as test ensemble. The former (heating) are clearly hcp-like, whereas the later resemble bcc-like structures.
The estimated values of $F_\mathrm{FP}-F_\mathrm{MLFF}$ are shown in Table~\ref{tab:TPT2}.
Obviously, MLFF-SVD is more accurate than MLFF-BLR for the free energies,
in particular for the bcc Zr where a larger deviation of $1.64$ meV/atom from the FP free energy is observed in the MLFF-BLR.
This is expected, since MLFF-SVD predicts more accurate potential energies as well as phonon dispersion relations.
For the free energy difference between the bcc and hcp phases, which is  relevant for estimating $T_c$,
MLFF-SVD and MLFF-BLR yield  deviations of 0.27 and 0.84 meV/atom, respectively, as compared to the one calculated by PBE.
After estimating the entropy difference between the two phases,
we estimate that this translates to an error of 9 K for MLFF-SVD in predicting $T_c$.
With the correction by TPT, our final estimate for $T_c$
by PBE is placed at 1049 K, in reasonable agreement with the experimental value of 1136~K.

\begin{table}
\caption {Estimated free energy difference $F_\mathrm{FP}-F_\mathrm{MLFF}$ (meV/atom) between FP and MLFF calculations
at $T$=1040 K using an ensemble of 40 structures picked from heating and cooling MD runs using MLFF-SVD.
Because of the hysteresis, the heating run yields hcp-like structures, whereas the cooling run yields
bcc-like structures.
}
\begin{ruledtabular}
\begin{tabular}{lccccc}
& \multicolumn{2}{c}{Heating/ hcp } & & \multicolumn{2}{c}{Cooling/ bcc} \\
 \cline{2-3}  \cline{5-6}
                  &  BLR  &  SVD   & &      BLR  &   SVD  \\
                  \hline
$F_\mathrm{FP}-F_\mathrm{MLFF}$ &  $-$0.80  &  $-$0.56  & &      $-$1.64 &   $-$0.83  \\
\end{tabular}
\end{ruledtabular}
\label{tab:TPT2}
\end{table}

\begin{figure}
\begin{center}
\includegraphics[width=0.5\textwidth,clip]{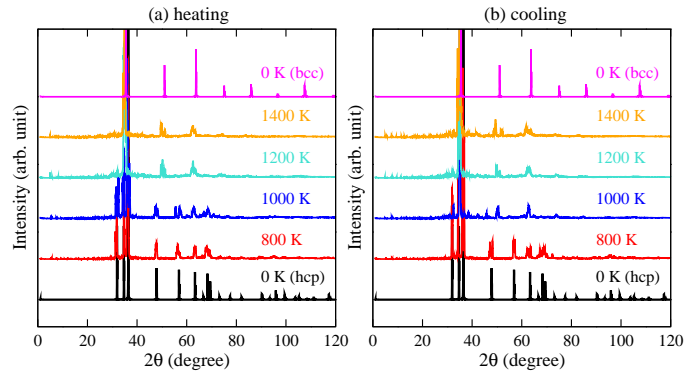}
\end{center}
 \caption{Simulated XRD patterns of Zr at selected temperatures during (a) heating  and (b) cooling MD simulations
 using MLFF-SVD. The corresponding adopted structures are shown in Fig.~\ref{fig:STR_MD_heat_cool}. The XRD patterns of
 hcp and bcc Zr at 0 K are also shown for comparison.
}
\label{fig:XRD_180}
\end{figure}

\begin{figure*}
\begin{center}
\includegraphics[width=0.8\textwidth,clip]{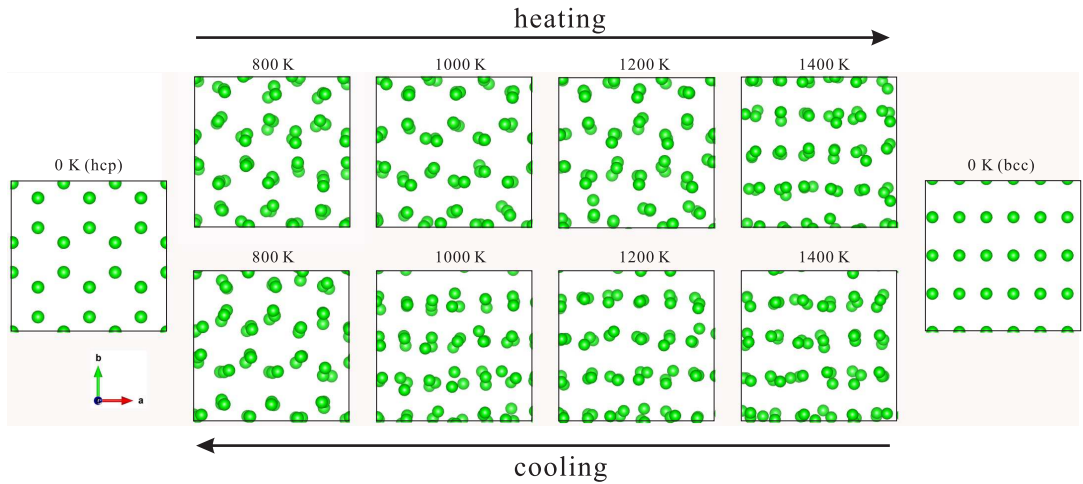}
\end{center}
 \caption{Structure evolution as a function of temperature during heating (upper row) and cooling (bottom row) MD simulations.
 These snapshot structures were picked from the MD trajectory using MLFF-SVD. The hcp and bcc structures
 at 0 K are also illustrated for comparison.
}
\label{fig:STR_MD_heat_cool}
\end{figure*}

To further validate that the observed phase transition is from hcp to bcc, x-ray powder diffraction (XRD) patterns are simulated for snapshot structures
picked from the MD trajectories. The results are shown in Fig.~\ref{fig:XRD_180}. From the XRD patterns, the
hcp-bcc phase transition is unambiguously confirmed, in accordance with Fig.~\ref{fig:Volume_vs_T}.
Furthermore, the displacive nature of the phase transition can be visually observed from the
changes in the atomic structure, as shown in Fig.~\ref{fig:STR_MD_heat_cool}.
The cooperative movement of Zr atoms of alternating (011)$_\beta$ planes
in the bcc phase along the opposite [01$\bar{1}$]$_\beta$ directions
results in the hcp atomic stacking sequence, confirming the the Burgers mechanism
for the temperature-driven bcc-hcp phase transition~\cite{BURGERS1934561}.

Our good prediction for the hcp-bcc phase transition of Zr undoubtedly demonstrates the strength and accuracy of on-the-fly MLFF.
In particular, almost no human interference was required during the training, which in the present study just involved heating and cooling of
hcp and bcc Zr. In principle, the training can be done in less than a week, with the human effort of setting up the calculations being just
few hours. As a matter of fact, testing the MLFF was a significantly more time-consuming endeavor in the present case.
Our MLFF training strategies and analysis presented in this work can also be employed
to study the temperature-dependent martensitic phase transitions in other materials such as other group-IV elements
Ti and Hf and  group-III elements Sc, Y, and La, with very little effort. In addition, the obtained force fields trained on hcp and bcc Zr
at ambient pressure can be further trained by applying external pressure and by including the
hexagonal $\omega$ phase in the training dataset so that the full temperature-pressure phase diagram of Zr
can be readily constructed.

\section{Conclusions}\label{sec:conlcusions}

To summarize, we have successfully applied the on-the-fly MLFF method to determine a force field for
bcc and hcp Zr and study the hcp-bcc phase transition of Zr. This is a fairly challenging problem
that is hard to address using brute force methods and  FP MD simulations due to the limited length and
time scales accessible to DFT simulations. Certainly, standard passive learning methods are possible
and have been successfully used in the past, but they do not offer the same sort of convenience
as the present approach.
The first-order displacive nature of the hcp-bcc phase transition---
manifested by an abrupt jump in the system volume and a change in the atomic stacking sequences
---has been unambiguously reproduced by our MD simulations
and identified by the simulated XRD patterns,
confirming the Burgers mechanism for the temperature-induced hcp-bcc phase transition.
In addition, our MLFF predicted phase transition temperature is found to be in reasonable agreement with experiment.
Finally, we have shown that due to the improved condition number,
SVD is in general more accurate than the regularized BLR,
which is evidenced by the systematic decrease of the errors in energies, forces, and stress tensors
for both the training and test datasets. The improvement by SVD over BLR
has also been showcased by its improved prediction of the energy difference between bcc and hcp Zr
and of the phonon dispersions of both hcp and bcc Zr. In summary, evidence shown in
this paper suggests that pseudo inversion of the design matrix using SVD is a
useful approach to overcome some of the limitations of regularized regression methods.

\begin{acknowledgments}
We thank E. Wimmer and J. Wormald for helpful discussions about this work.
This work was funded by the Advanced Materials Simulation Engineering Tool (AMSET) project, sponsored by the US Naval Nuclear Laboratory (NNL) and directed by Materials Design, Inc.
\end{acknowledgments}

\bibliographystyle{apsrev4-1}
\bibliography{reference} 

\end{document}